\documentclass[aps,prl,twocolumn,a4paper,10pt,notitlepage,footinbib,superscriptaddress,showpacs]{revtex4-1}%
\usepackage[english]{babel}
\usepackage[T1]{fontenc}
\usepackage{endnotes}
\usepackage{amssymb,amsmath,amsfonts}
\usepackage{textcomp}
\usepackage{graphicx,color}
\usepackage[utf8x]{inputenc}
\usepackage{float}
\usepackage{placeins}
\usepackage{multirow}
\usepackage{colortbl}
\usepackage{tabulary}
\usepackage{etoolbox}

\begin{document}

\title{Activity induced isotropic-polar transition in active liquid crystals}

\author{Maria Grazia Giordano}
\affiliation{Dipartimento  di  Fisica,  Universit\`a  degli  Studi  di  Bari  and  INFN,  via  Amendola  173,  Bari,  I-70126,  Italy}

\author{Francesco Bonelli}
\affiliation{Dipartimento di Meccanica, Matematica e Management, DMMM, Politecnico di Bari, 70125, Bari, Italy}

\author{Livio Nicola Carenza}
\affiliation{Dipartimento  di  Fisica,  Universit\`a  degli  Studi  di  Bari  and  INFN,  via  Amendola  173,  Bari,  I-70126,  Italy}

\author{Giuseppe Gonnella} 
\affiliation{Dipartimento  di  Fisica,  Universit\`a  degli  Studi  di  Bari  and  INFN,  via  Amendola  173,  Bari,  I-70126,  Italy}

\author{Giuseppe Negro}
\affiliation{Dipartimento  di  Fisica,  Universit\`a  degli  Studi  di  Bari  and  INFN,  via  Amendola  173,  Bari,  I-70126,  Italy}

\title{Activity induced isotropic-polar transition in active liquid crystals}

\begin{abstract}
Active fluids are intrinsically out-of-equilibrium systems due to the
internal energy injection of the active constituents. We show here that 
a transition from a motion-less isotropic state towards a flowing polar 
one can be possibly driven by the sole active injection through the 
action of polar-hydrodynamic interactions in absence of an \emph{ad hoc} free-energy 
which favors the development of an ordered phase.
In particular, we  propose  an  analytical  argument and we perform 
lattice Boltzmann simulations where the appearance of large 
temporal fluctuations in the polar fraction of the system is observed 
at the transition point.
Moreover, we make use of a scale-to-scale analysis to unveil the energy transfer mechanism,
proving that elastic absorption plays a relevant role in the overall dynamics of the system,
contrary to what reported in previous works on the usual active gel theory where this term 
could be factually neglected.
\end{abstract}

\maketitle

\section{Introduction}

Active fluids intrinsically evolve out of equilibrium due to internal energy injection~\cite{ramaswamy2010,marchetti2013}.
Systems of biological origin -- such as bacterial~\cite{wensink2012,Bratanov2015} and cytoskeletal~\cite{bendix2008,Decamp2015,Guillamat2017} suspensions -- or synthetic realizations, as Janus particles~\cite{ebbens2014} and polyacrylic acid hydrogels~\cite{korevaar2020}, convert chemical energy stored in some \emph{reservoir} into \textcolor{black}{non-equilibrium} stresses which result in self-sustained flows~\cite{Wioland2013,Decamp2015,dunkel2013}. 
Recent research has proved that the mutual interactions between hydrodynamics and active energy input 
may give rise to non-equilibrium dynamical states connected to a plethora of unexpected behaviors as, for instance, super-fluidic and negative viscosity states~\cite{hatwalne2004,Liverpool2006,PhysRevLett.104.098102,Guo201722505,negro2019,gachelin2013,lopez2015}, spontaneous flows~\cite{sankararaman2009,voituriez2005,Keber1135,Carenza22065,carenza2020_physA,giomi2008,tjhung2012,giomi2014,bone2017} and active turbulence regime~\cite{wolgemuth2008,giomi2015,wensink2012,Urzay2017,Alert2020,carenza2020}.

Even with intrinsic structural differences, a common feature to many active systems  is the breaking of isotropic symmetry. This may either occur at the level of the individual constituents~\cite{sanchez2012,nedelec1997} -- since their bodies often exhibit an elongated or filamentous structure -- or due to the direction of motion~\cite{saintillan2007,wensink2012,vics1995,baskaran2008}, resulting into 
the emergence of local patterns with polar/nematic or even hexatic symmetry~\cite{cugliandolo2017} characterized by the proliferation of topological defects whose behavior strongly differs from their passive counter-part~\cite{Decamp2015,Thot2002,Zhou2014,putzig2016,giomi2013,Elgeti2011,Doostmohammadi2017,negro2018}.

However, experiments on concentrated suspensions of bacteria and microtubules show that polar and nematic order \emph{only} emerges in presence of activity, while isotropic symmetry is restored in the passive limit~\cite{Opathalage4788,sanchez2012,Lushi9733}, thus suggesting that the thermodynamic \emph{ground state} of a \emph{passive} suspension is basically disordered~\cite{Fodor2016} and activity-induced visco-elastic couplings may alone drive a transition to an ordered state
\textcolor{black}{by effective aligning interactions among the individual constituents.}
This is somehow in contrast with the typical approach of many models for active fluids -- as the well-known 
active gel theory~\cite{marchetti2006,carenza2019} -- where the existence of an equilibrium free-energy favouring an ordered equilibrium configuration in absence of activity is postulated \emph{a priori}~\cite{kruse2004,Cates2008,giomi2015,blow2014,Shendruk2017}.

The physical origin of such motility-induced transition from the isotropic to an ordered state is to be related to the intricate nemato-hydrodynamic behavior and the connected rheological properties of liquid-crystal suspensions. For instance, Markovich~\emph{et al.}~\cite{Markovich2019} have recently shown, analytically and numerically, that an external shear can induce a first-order phase transition from the isotropic to the ordered state in a (passive) polar liquid crystal. The mechanism at base of this effect lies in the renormalization of the bulk constants describing the ordering properties of the polarization, due to the strain effect of the mechanically imposed shear flow, thus allowing for the development of a polar phase in a region of the parameter space which would otherwise be isotropic.
This is of intereset for our work, since the swimming mechanism of active liquid crystals is able to generate local shear flows which may possibly alter the threshold of the orientational transition. Indeed, extensile swimmers are able to expel fluid at their ends and  draw it across their body acting on the surrounding environment as out-warding force dipoles, while contractile swimmers act as in-warding force dipoles, leading to a reversed pattern~\cite{hatwalne2004}. In both cases, such active flow circulation  is able to substantially affect the overall dynamics of the system~\cite{Cates2008,Giomi2010} and its rheological properties~\cite{negro2019,lopez2015,Guo201722505,carenza2020_sci_rep}.

Our main objective in this Letter is to investigate the possibility that the flows internally generated by polar active constituents may induce orientational order even in absence of an \emph{ad hoc} free-energy, analogously to what happens in the case of an external shear.
A similar study has been carried out by Thampi~\emph{et al.}~\cite{Thampi2015} for the case of a system with nematic symmetry, proving that activity alone is actually able to  affect the degree of nematic ordering. 
They found flow-driven macroscopic fluctuations of the concentration and phase-separation, resulting into a dynamical mixture of extensile and contractile regions.

We will show that an isotropic-ordered phase transition also develops in polar systems, exhibiting yet a different nature due to different symmetry features.
By means of numerical simulations, we will show that activity is able to give rise to a first-order isotropic-polar (I-P) transition for extensile arrow-like systems and for contractile disk-like ones, in presence of nemato-hydrodynamic interactions. 
In addition, we perform a scale-to-scale analysis in Fourier space to investigate the intricate energy-transfer mechanism in our model, finding that the most relevant contribution to energy absorption is the polar one, contrary to what reported in analogous investigations~\cite{giomi2015,carenza2020_bif,Urzay2017,Alert2020} on usual active gel theory.

\section{Model} 
We consider a polar fluid in a $2d$ geometry with mass density $\rho$ and velocity $\bm{v}$. The ordering properties of
the suspended particles are encoded in the polarization field $\bm{P}$, while the concentration of the nutrient is denoted with $\phi$.
The equations which rules the hydrodynamic of the system are
\begin{align}
\nabla \cdot \bm{v} = 0 \label{eqn:continuity}\\
\rho (\partial_t \bm{v} + \bm{v} \cdot \nabla \bm{v}) = -\nabla p + \nabla \cdot \bm{\sigma}  \label{eqn:NSE}
\end{align}
where the first one is the solenoidal condition for the flow field, enforcing the incompressibility of the system, 
and the second one is the Navier-Stokes equation. Here, $p$ is the hydrodynamic pressure, while $\bm{\sigma}$ is the stress tensor.
This is in turn the sum of three contributions. The first is the usual viscous stress
$$
\sigma^{visc}_{\alpha\beta} = \eta (\partial_\alpha v_\beta + \partial_\beta v_\alpha),
$$
where $\eta$ is the nominal viscosity of the fluid and it is responsible for
energy dissipation. The polar stress~\cite{degennes1993} 
\begin{eqnarray*}
\sigma_{\alpha\beta}^{pol}=\frac{1}{2}(P_{\alpha}h_{\beta} -P_{\beta}h_{\alpha})-\frac{\xi}{2}(P_{\alpha}h_{\beta}+P_{\beta}h_{\alpha})\nonumber\\
- k_{P} \partial_{\alpha}P_{\gamma}\partial_{\beta}P_{\gamma}\label{eq:elastic-stress},
\end{eqnarray*}
provides a non-linear coupling between the dynamics
of the fluid and the evolution of the polarization field. The explicit structure of the molecular field $\bm{h}$ and the role of the flow-aligning parameter $\xi$ will be discussed in the following. 
Finally, the active stress tensor
$$
\sigma^{act}_{\alpha \beta} = - \zeta \phi \left( P_\alpha P_\beta -\dfrac{|\mathbf{P}|^2}{2} \delta_{\alpha \beta} \right)    
$$
is a phenomenological term~\cite{simha2002} responsible for the local energy input due to the active constituents whose rate is proportional to the activity parameter $\zeta$ --positive for extensile swimmers and negative for contractile ones-- and to the nutrient concentration $\phi$  which in turn evolves through an advection-diffusion equation
\begin{equation}
\partial_t \phi + \bm{v} \cdot \nabla \phi = D \nabla^2 \phi,
\label{eqn:diffusion}
\end{equation}
where $D$ is the diffusion constant.

\begin{figure}[bt]
\centering
\includegraphics[width=1.0\columnwidth]{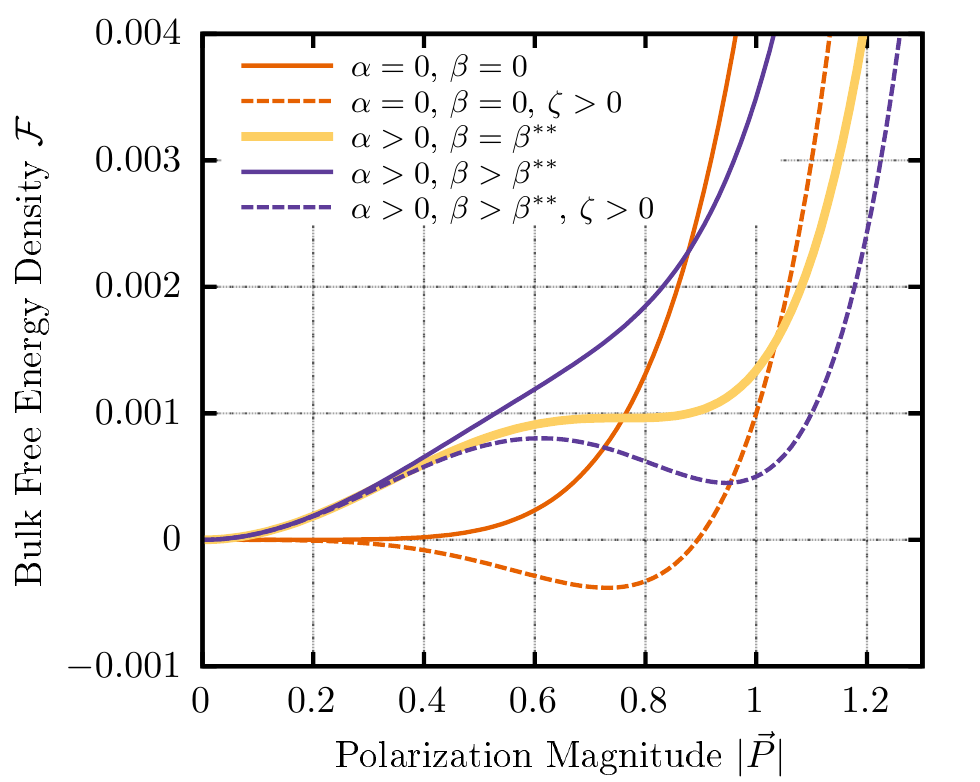}
\caption{\textbf{Effective free energy in active polar gels.} Continuous lines show the functional dependence of the free energy from the polarization field. For positive $\alpha$ and $\gamma$ and $\beta>\beta^{**}$, the only stable solution is $\bm{P}=0$ corresponding to the isotropic phase, while at $\beta=\beta^{**}$ the free-energy exhibit an inflection point at $\bm{P}\neq 0$. Dashed lines represent the corresponding free energies after the renormalization of the coefficient of the quartic term $\beta \longrightarrow \Tilde{\beta}$, in accordance to Eq.~\eqref{eqn:renormalization}.}
\label{fig1}
\end{figure}

The evolution equation for the polarization field is given by the Ericksen-Leslie equation~\cite{degennes1993,marchetti2013} adapted for the treatment
of a vector order parameter and reads as follows
\begin{equation}
d_t \bm{P} = \xi\bm{D}\cdot\bm{P} +\frac{1}{\Gamma} \bm{h},
\label{eqn:P_eq}
\end{equation}
where we denoted with $d_t \bm{P}$ the sum of the material derivative $\partial_t \bm{P}+\left(\bm{v}\cdot\nabla\right)\bm{P}$ and the co-rotational derivative $ \bm{\Omega}\cdot\bm{P}$.
Here, $D_{\alpha \beta}=(\partial_\beta v_\alpha + \partial_\alpha v_\beta)/2$ and $\bm{\Omega}_{\alpha \beta}=(\partial_\beta v_\alpha - \partial_\alpha v_\beta)/2$ are respectively the symmetric and anti-symmetric parts of the strain tensor $\nabla \bm{v}$. The alignment parameter $\xi$ controls the aspect ratio of the suspended particles ($\xi>0$ for rod-like particles and $\xi<0$ for disk-line ones) as well as the response to a shear flow. In particular, when $|\xi|>1$, the polarization field  uniformly orients at the Leslie angle with respect to the imposed flow~\cite{Markovich2019,Larson1999}. In this case, the liquid crystal is said to be \emph{flow-aligning}. Conversely, when $|\xi|<1$, the liquid crystal rotates under the effect of shear and is customarily  addressed as \emph{flow tumbling}.
The rotational viscosity $\Gamma$ measures the relevance of the hydrodynamic interactions with respect to the driving effect of the molecular field $\bm{h}=-\delta \mathcal{F}/\delta \bm{P}$,  where $\mathcal{F}$ is a suitable free energy which we introduce to guarantee the thermodynamic stability of the system. To do so, we consider
an expansion in power of $|\bm{P}|^2$ up to sixth order:
\begin{equation}
\mathcal{F}= \int \textit{d}\bm{r} \left[ \alpha |\bm{P}|^2 + \beta |\bm{P}|^4 + \gamma |\bm{P}|^6 + \dfrac{k_P}{2} (\nabla \bm{P})^2 \right]
\label{eqn:freeE}
\end{equation}
where $k_P$ is the elastic constant of the polar phase and $\alpha$, $\beta$ and $\gamma$ are three bulk phenomenological parameters with $\gamma>0$ to ensure stability.
When both $\alpha$ and $\beta$ are positive or null (see continuous orange line in Fig.~\ref{fig1}, showing the possible behaviors of the bulk potential relevant for the following discussion), the only stable solution is the trivial one at $\bm{P}=0$, corresponding to the isotropic phase. This condition remains unaltered even for positive values of $\alpha$, as long as $\beta>\beta^{**}=-\sqrt{3\alpha \gamma}$ (see continuous purple line in Fig.~\ref{fig1}), where $\beta^{**}$ is a critical value corresponding to the appearance of a meta-stable ordered region  at $\bm{P}\neq 0$ (yellow line in Fig.~\ref{fig1}). \textcolor{black}{Finally for $\beta<\beta_{cr}=-\sqrt{4 \alpha \gamma}$ the minimum corresponding to the polar phase becomes absolute.}

\textcolor{black}{In the usual active gel theory, the parameters are usually chosen to ensure the free-energy to have a global minimum corresponding to non-null polarization. For the case of a polar gel, this can be obtained by making use of a double-well potential by setting $\alpha<0$,  $\beta>0$ and $\gamma=0$~\cite{kruse2004}.}
Conversely, here we will only consider cases with $\alpha>0$ and $\beta>\beta^{**}$, whose equilibrium configuration is the isotropic one. However, it turns out that visco-elastic interactions arising due to activity-induced flows, may lead to a significant renormalization of the bulk coefficients, leading to a situation  where polar order is also observed. 
\textcolor{black}{To show this, we start from the assumption --which we will verify \emph{a posteriori} in numerical simulations-- of local mechanical equilibrium $\bm{\sigma}^{visc}+\bm{\sigma}^{pol}+\bm{\sigma}^{act} \approx 0$. 
Retaining only the lowest bulk order contributions to the molecular field ($\sim \bm{P}$) allows us to find a functional relation between the flow structure and the polarization
\begin{equation}
D_{\alpha \beta} \approx \dfrac{\zeta \phi}{2 \eta }  \left(P_\alpha P_\beta - \dfrac{|\bm{P}|^2}{2} \delta_{\alpha \beta} \right) - \dfrac{\alpha\xi}{\eta} P_\alpha P_\beta .
\label{eqn:P_flow}
\end{equation}
By substituting this relation into Eq.~\eqref{eqn:P_eq} we obtain:
\begin{equation}
d_t \bm{P} = \dfrac{\xi}{\eta} \left(\dfrac{\zeta \phi}{4} - \alpha \xi \right) |\bm{P}|^2 \bm{P}  +\frac{1}{\Gamma} \bm{h}.
\end{equation}
This new term has cubic dependence and induce a renormalization of the coefficient of the quartic term in Eq.~\eqref{eqn:freeE} as follows
\begin{equation}
\beta \longrightarrow \Tilde{\beta} = \beta - \dfrac{ \xi \Gamma}{4 \eta } \left( \dfrac{\zeta \phi}{4} - \alpha \xi \right),
\label{eqn:renormalization}
\end{equation}
so that for strong enough active forcing 
\begin{equation}
\zeta \xi >  \dfrac{4 \alpha \xi^2}{\phi} + \dfrac{16  \eta}{\phi \Gamma} \left(\beta + \sqrt{\Theta \alpha \gamma} \right),
\label{eqn:threshold}
\end{equation}
the renormalized free energy may either develop a local minimum  at $\bm{P} \neq 0$, (for $\Theta=3$ corresponding to the critical condition $\tilde{\beta}<\beta^{**}$) or a global one (for $\Theta=4$ corresponding to  $\tilde{\beta}<\beta_{cr}$),
as suggested by the dashed lines in Fig.~\ref{fig1}.}
This may either occur for extensile rod-like particles or contractile disk-like ones.
In particular, we observe that the predicted nature of the transition is first order, hence in the range $\beta_{cr} < \tilde{\beta}(\zeta,\xi)< \beta^{**}$ we expect phase coexistence of polar (active) and isotropic (passive) regions. The transition becomes second order in the limit $\alpha, \beta=0$.

\section{Uniform nutrient concentration}
We numerically integrate the hydrodynamic equations Eqs.~\eqref{eqn:continuity}-\eqref{eqn:NSE} on a squared computational grid of size $L=256$  by means of a predictor-corrector LB model with periodic boundary conditions. The dynamics of the order parameter $\phi$ and $\bm{P}$ is solved through a predictor-corrector finite-difference algorithm, implementing first-order upwind scheme and fourth order accurate stencils for space derivatives~\cite{denniston2001,carenza2019}.
Except otherwise stated, we set $\eta=5/3, D=4 \times 10^{-4}$ and $\xi=1.1$, while the polarization field is initialized with random orientation and $|\bm{P}|=1$.
Moreover we choose $\alpha=\gamma=0.005$ $\beta=-0.0065>\beta^{**}$, deep in the isotropic region of the parameter space (see Fig.~\ref{fig1}, continuous purple line).
\begin{figure}[t]
\centering
\includegraphics[width=1.0\columnwidth]{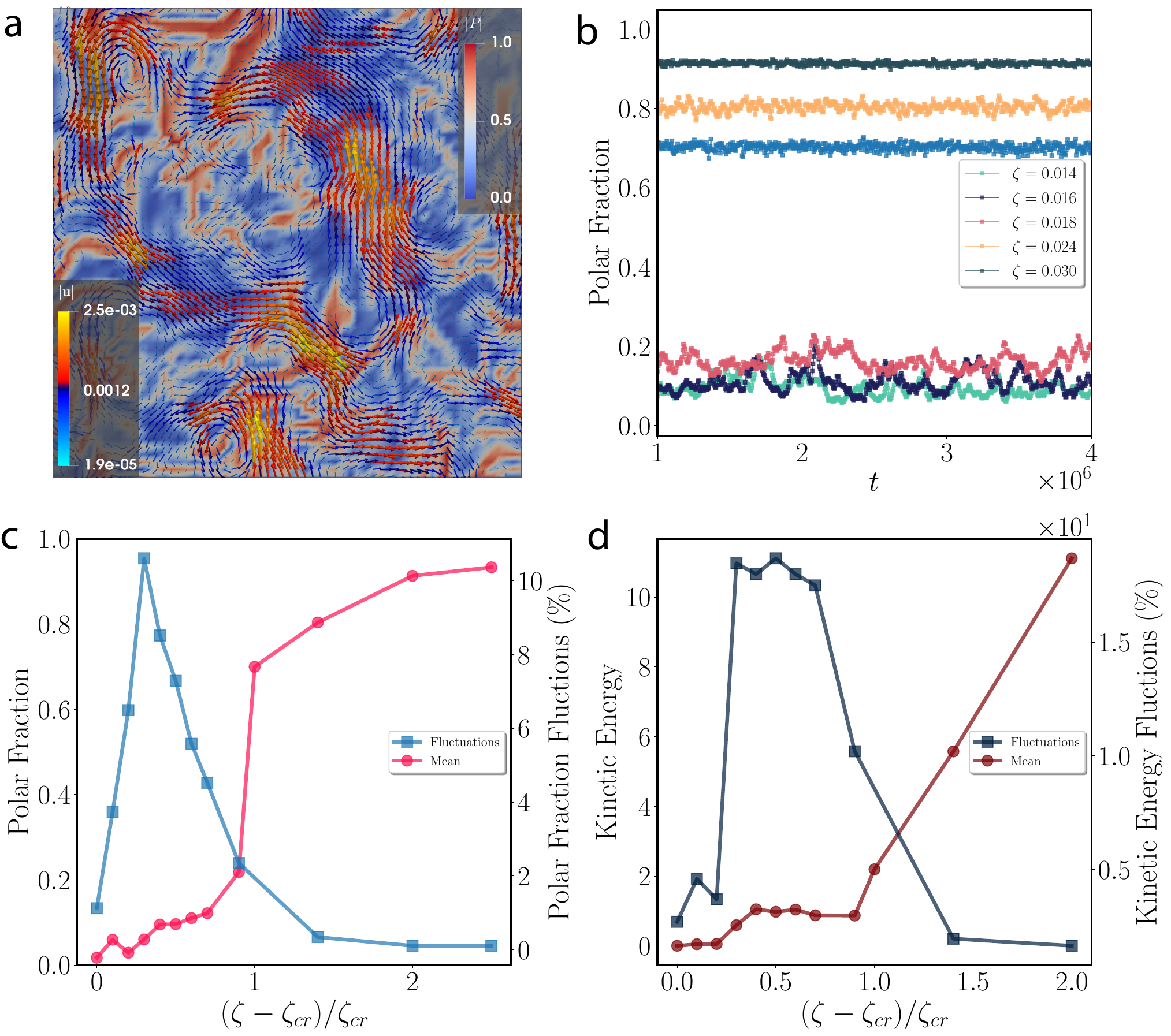}
\caption{\textbf{Uniform nutrient concentration.} (a) 
Velocity field superimposed to the contour plot of the magnitude of $\bm{P}$ on a $64^2$ fraction of the system for $\xi=1.1$ and $\zeta=0.015$ (($\zeta- \zeta_{cr})/\zeta_{cr}=0.5$). 
(b) Polar fraction temporal evolution for some values of the activity $\zeta$.
Panels~(c) and~(d) respectively show the behavior of the polar fraction and kinetic energy and their fluctuations, at varying $\zeta$ across the isotropic-polar transition ($\zeta_{cr}=0.01$). Temportal fluctuations are computed here and elsewhere as percent deviation from the mean.}
\label{fig2}
\end{figure}

We first consider the case of an active polar gel with uniform nutrient concentration (\emph{i.e.} $\phi=1$ in the whole volume and constant in time).
For $0<\zeta \lesssim \zeta_{cr}=0.01$, the rate of active injection is not strong enough to affect the relaxation towards the ground state and the polarization rapidly disappears, leaving the system in an isotropic motion-less state.

As activity is increased over the critical threshold $\zeta_{cr}$, a dynamical non equilibrium state sets up, characterized by the coexistence of disordered/passive regions and polar/active domains with elongated shape, where $\bm{P}$ is consistently different from $0$ (see for instance panels~(a) and~(b) of Fig.~\ref{fig2}). These evolve in time under the fueling action of the activity and continuously shrink and enlarge, giving rise to large fluctuations of the mean polar fraction $\psi_P$ of the system (Fig.~\ref{fig2}(c)) computed as
$$
\psi_P(t)= \dfrac{1}{L^2} \langle \int  \vartheta(|\bm{P}|-\bar{P}) \textit{d}\bm{r} \rangle
$$
where $\vartheta(\cdot)$ is the Heaviside step function and $\bar{P}=0.2$ is a suitable threshold to distinguish polar regions from isotropic ones and $\langle \dots \rangle$ denotes the time average.
More specifically, regions of well aligned liquid crystal undergo a banding instability. In the usual active gel theory this mechanism is at the the base of the production of topological defects ~\cite{giomi2013,Doostmohammadi2017}. 
However, in our model, strong distortions of the polarization pattern are not sustainable since they are associated with a reduction of the polar order, which results in the thinning of the polar domains and eventually in their disappearance, due to the diffusive elastic term $\sim -k_P \nabla^2 \mathbf{P}$ in the molecular field $\bm{h}$.

\begin{figure}
\centering 
\includegraphics[width=1.0\columnwidth]{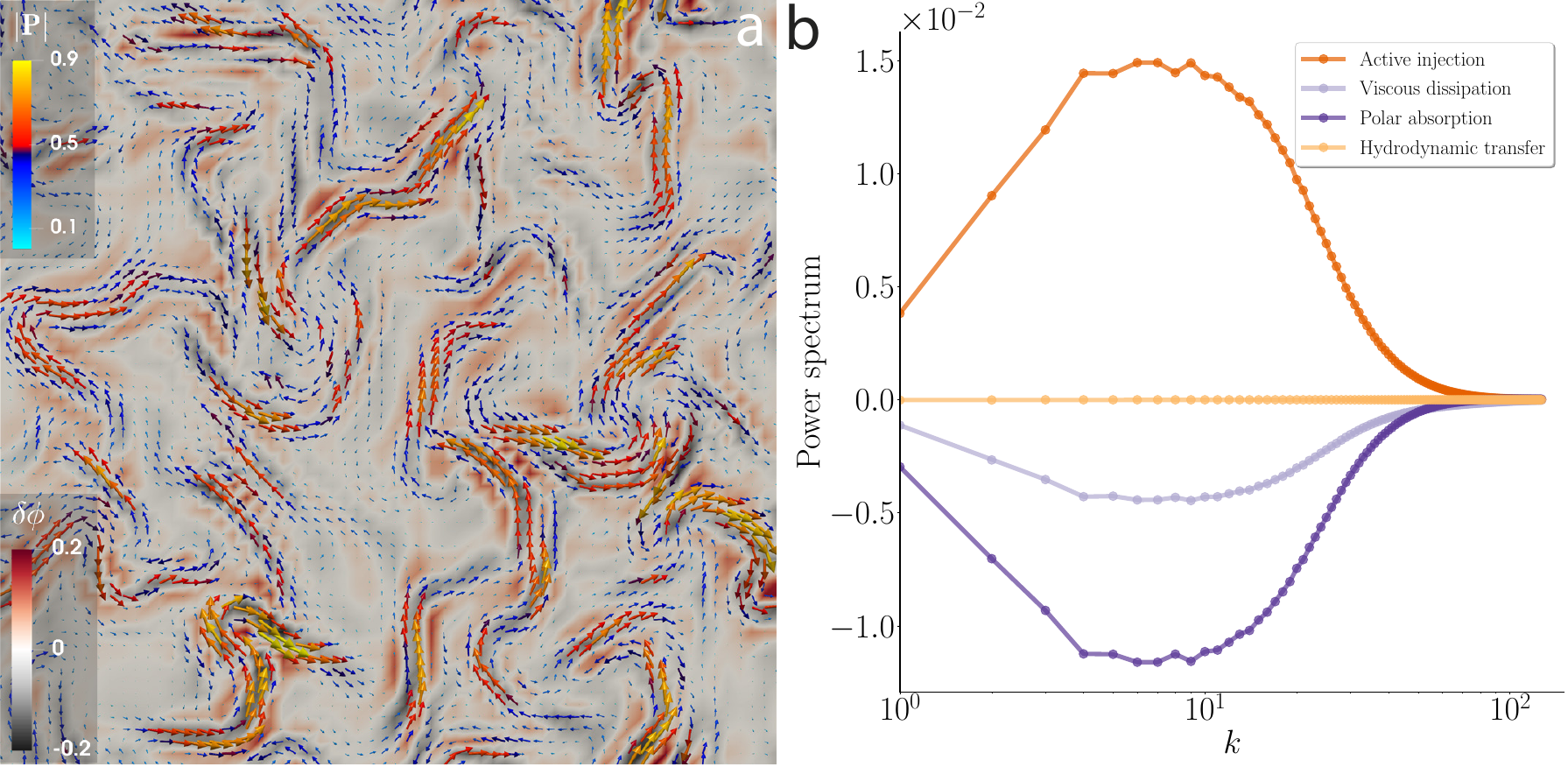}
\caption{\textbf{Dynamical nutrient concentration.}  Panel~(a) shows a typical non-equilibrium steady-state configuration of the polarization field (colored vectors) superimposed to the contour plot of the concentration fluctuations $\delta \phi=\phi-\phi_0$, for $\xi=1.1$ and $\zeta=0.015$. Notice that polar/isotropic regions correspond to negative/positive values of $\delta\phi$. Panel~(b) shows that the energy balance in Fourier space (see Eq.~\eqref{eqn:energy}) satisfies the hypothesis of local mechanical equilibrium. }
\label{fig3}
\end{figure}

Nevertheless, these regions of deformation also act as a source of momentum --since the active force  $\bm{f}^{act} \propto \nabla \cdot \bm{P}\bm{P}$-- and produce jets of fluid which advect and shear the liquid crystal, counteracting elastic effects and reinforcing the polar order.
This mechanism rapidly becomes more and more dominant as extensile activity is increased over $2\zeta_{cr}=0.02$. Activity-induced flows strengthen, causing a sharp increase of the total kinetic energy as shown in Fig.~\ref{fig2}(d), and invade the whole computational domain, pushing the system into a polar state with $\psi_P \gtrsim 0.9$ and negligible fluctuations.
This state, characterized by vortical regions both in the velocity and the polarization field, can be considered the analogue of the \emph{active turbulent} regime for the usual active gel theory.

Hence, our results suggest that activity alone is able to drive a transition  from a quiescent isotropic state at low activity towards a dynamical polar state where rotational symmetry is restored at a statistical level due to the chaotic vortical patterning of the polarization field.
Such transition develops in an extended range of the activity parameters ($0.01 \lesssim \zeta \lesssim 0.02$) and exhibits the typical features of a first-order transition, including coexistence between ordered and disordered regions and large fluctuations. These values are of the same order of magnitude and roughly in agreement with those obtained through Eq.~\eqref{eqn:threshold} which predicts the polar metastability region in the range $0.020 \lesssim \zeta \lesssim 0.022$ ($1<(\zeta-\zeta_{cr})/\zeta_{cr} < 1.2$). 
A further increment of the activity eventually leads the system in a region of metastability for the isotropic phase where the effective minimum for $\bm{P}$ is favored and the polar fraction saturates.

\section{Dynamical nutrient concentration}
\begin{figure}[t]
\centering
\includegraphics[width=1\columnwidth]{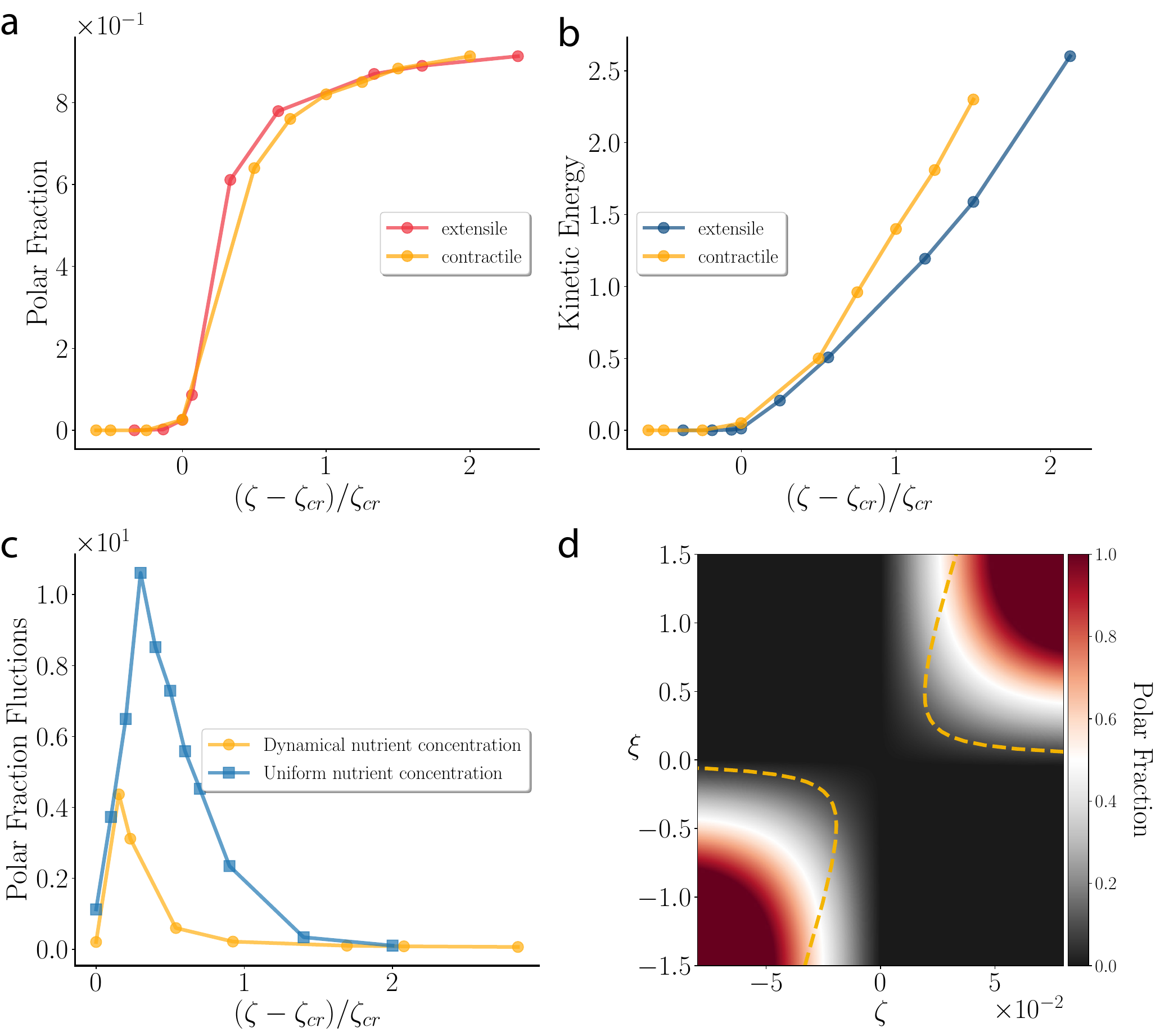}
\caption{\textbf{Polar phase diagram.} Panels~(a) and (b) respectively show the behavior of the polar fraction and the mean kinetic energy at varying activity for an extensile system with $\xi=1.1$ and a contractile one with $\xi=-1.1$. The critical extensile (contractile) threshold is $\zeta_{cr}=0.012 \ (0.015)$. Panel~(c) shows the comparison of the time fluctuations of the polar fraction between the case with uniform and dynamical nutrient concentration. Panel~(d) shows the contour plot of the polar fraction in the $\xi-\zeta$ plane. The dashed yellow line represent the theoretically predicted transition line of  Eq.~\eqref{eqn:threshold}.}
\label{fig4}
\end{figure}

We shall now treat the concentration of the nutrient $\phi$  as a dynamical field and let it evolve from its uniform initial state ($\langle \phi\rangle$=1) according to Eq.~\eqref{eqn:diffusion}.
T
For small active doping the system ends up in an isotropic quiescent state, while for $\zeta>\zeta_{cr}=0.012$ fluctuating polar domains begin to populate the system. Their dynamics is accompanied by  fluctuations in the concentration field caused by the advective effect of active flows. Interestingly, $\phi$ tends to lower in those regions populated by the polarization (see Fig.~\ref{fig3}). This can be easily understood in terms of the relation between $\bm{P}$ and the flow field. Indeed Eq.~\eqref{eqn:P_flow} suggests that $\partial_\alpha v_\beta \sim (\zeta \phi - 2 \alpha \xi) P_\alpha P_\beta / \eta$ hence, when the condition in Eq.~\eqref{eqn:threshold} is fulfilled, we get from Eq.~\eqref{eqn:diffusion} that $\partial_t \phi \sim - \phi \bm{P}^2<0$. 
Therefore, the concentration field is advected towards isotropic regions, leading to an effective increment in the rate of active injection which favors in turn the formation of polar regions. This is at the base of the observed $\phi$ and $\bm{P}$ patterns in ours simulations, where thin stripes of polarization are interspaced with isotropic regions (see Fig.~\ref{fig3}a).
The presence of a dynamical concentration field has the important effect to reduce the temporal fluctuations in the polar fraction $\psi_P$ at the isotropic/polar transition (see comparison in Fig.~\ref{fig4}c) which now develops in a more narrow range of activities, signalled by a sharp increase of $\psi_P$ in correspondence of the critical activity $\zeta_{cr}$ (see red lines in Fig.~\ref{fig2}c and~Fig.\ref{fig4}a).

A feature that can enlighten the dynamical role of polarization comes from the analysis of different contributions to energy transfer. We shall now consider
a scale-to-scale balance equation for the kinetic energy in Fourier space:
\begin{equation}
\partial_t E_k + \mathcal{T}_k = \mathcal{S}^{visc}_k + \mathcal{S}^{pol}_k + \mathcal{S}^{act}_k,
\label{eqn:energy}
\end{equation}
where the energy spectrum  $E_k=\langle  | \mathbf{v}_\mathbf{k}|^2 \rangle_\Omega /2$ and $\langle \dots \rangle_\Omega$ stands for the average in Fourier space over shells of equal momentum ($|\mathbf{k}|=k$). $\mathcal{T}_k=  \langle \mathbf{v}_\mathbf{k}^* \cdot \mathbf{J}_\mathbf{k} \rangle_\Omega$ represents the advective transfer rate, being $\mathbf{J}_\mathbf{k}$ the hydrodynamic flux $-\nabla p + \mathbf{v}\cdot \nabla \mathbf{v}$ in the $k$-space.
The terms $\mathcal{S}_k$ are source/sink terms for the kinetic energy. In particular, 
$$
\mathcal{S}^{visc}_k= 2 \pi i \langle \mathbf{v}_\mathbf{k}^* \otimes \mathbf{k} : \sigma^{visc}_\mathbf{k} \rangle_\Omega/L = - \frac{8 \pi^2 \eta}{L^2}  k^2 E_k
$$
gives the rate at which energy is dissipated by viscosity and 
$$
\mathcal{S}^{pol}_k= 2 \pi i \langle \mathbf{v}_\mathbf{k}^* \otimes \mathbf{k} : \sigma^{pol}_\mathbf{k} \rangle_\Omega/L 
$$
gives the rate at which the polarization field absorbs energy to be maintained in its non-equilibrium state. Finally, 
$$
\mathcal{S}^{act}_k= 2 \pi i \langle \mathbf{v}_\mathbf{k}^* \otimes \mathbf{k} : \sigma^{act}_\mathbf{k} \rangle_\Omega/L 
$$
gives the rate at which energy is injected in the system by the activity.

Panel~(b) of Fig.~\ref{fig3} shows the energy balance for the case at $\zeta=0.03$.
As expected, the active power spectrum injects energy over all scales ($\mathcal{S}^{act}_k>0$) and is counter-balanced by viscous dissipation and polarization absorption, while the non-linear hydrodynamic advective term $\mathcal{T}_k \approx 0$, in accordance to the value of the Reynolds number $Re = \rho l^* v^*/\eta$ (where $l^*$ and $v^*$ are respectively a typical scale and velocity of the flow~\cite{carenza2020_bif}) which never exceeds $10^{-1}$ in our simulations.
We found that the polar term gives here a greater contribution to energy absorption than viscosity. This is in contrast with respect to previous studies which accounted for the energy transfer in active systems, where the polar term  could be factually neglected~\cite{giomi2015,carenza2020_bif,Urzay2017,Alert2020}.
The explanation lies in the fact that while in the usual active gel approach, the polarization field can be assumed close to equilibrium with the only relevant contribution arising from  deformations ($\bm{h} \sim - k_P \nabla^2 \bm{P}$) in the present case activity must provide enough energy to overcome the energy barrier due to bulk terms, which would drive the system into an isotropic state in absence of active forcing.

Power spectra behavior exhibits a localized balance in Fourier space, in agreement with our assumption of local mechanical equilibrium. Moreover, the most relevant scales $\lambda$, at which the rate of active injection reaches its maximum for the case in Fig.~\ref{fig3}, range in $64 < \lambda < 13$ (with $\lambda=L/k$) corresponding to the typical size of the structures developed by the polarization field.

Finally, we comment on the effect of the alignment parameter $\xi$. At $\xi=0$ we do not observe a polar state for any value of activity, proving that the alignment term $\sim \bm{D} \cdot \bm{P}$ is crucial to the development of the nemato-hydrodynamic instability discussed in this paper.
By considering progressively higher values of $|\xi|$ we find that the transition towards the ordered state develops at smaller and smaller values of $|\zeta|$.
This mechanism drastically depends on the sign of both $\xi$ and $\zeta$, as suggested by Eq.~\eqref{eqn:threshold}. Indeed, the phase-diagram in Fig.~\ref{fig4}d shows that polar regions are only observed for either extensile rod-like particles ($\zeta>0,\xi>0$) or contractile disk-like ones ($\zeta<0,\xi<0$). The observed position of the transition line in the $\xi-\zeta$ plane is in good agreement with the theoretical prediction of Eq.~\eqref{eqn:threshold} (see dashed line in Fig.~\ref{fig4}d).
Moreover, the response of the system is roughly symmetric under a sign change in both the aligning parameter and the activity, as suggested by the comparisons between the polar fraction and the kinetic energy for the extensile and contractile case at $|\xi|=1.1$ respectively shown in Fig.~\ref{fig4}(a) and (b).

\section{Conclusion}
In this Letter we have investigated the dynamics of an active polar gel in absence of a free-energy favoring a polar state, contrary to most previous work based on active gel theory where an ordered phase is obtained by opportunely setting the parameters of some free-energy functional.
We found that nemato-hydrodynamic couplings may induce a transition from the motion-less and energetically favored disordered phase towards a self-sustained flowing polar state evolving far from equilibrium.
In our model the rate of energy absorption due to the polarization cannot be neglected --as we showed through a scale-to-scale analysis in Fourier space-- contrary to what reported by previous works on energy transfer in active gels. 
On these assumptions, we proposed an analytical argument which suggests such transition to be first-order. This picture was confirmed 
through numerical experiments which showed that coexistence of polar and isotropic regions occurs over a large range of activities at the transition, accompanied by important temporal fluctuations in the polar fraction of the system.
Moreover, in presence of a diffusive field describing the concentration of the nutrient, the transition becomes sharper and fluctuations are considerably reduced. 


We believe that results of our paper  provide a solid ground for a deeper understanding  of the motility of active gels and furnish a starting point to study the effects of confinement. In this case, the presence of boundaries either soft or rigid may result into dynamical states which are commonly observed in experiments~\cite{sanchez2012}.

We warmly thank Julia Yeomans and Davide Marenduzzo for discussions from which this work originated.
The work has been performed under the Project HPC- EUROPA3 (INFRAIA-2016-1-730897), with the support of the EC Research Innovation Action under the H2020 Programme; in particular, the authors gratefully ac- knowledges the support and the computer resources pro- vided by EPCC.

\bibliographystyle{unsrt}
\bibliography{biblio}

\end{document}